\documentclass[preprint,showpacs,preprintnumbers,amsmath,amssymb]{revtex4}

\usepackage{graphicx}
\usepackage{dcolumn}
\usepackage{bm}
\usepackage{multirow}

\begin{document}

\preprint{APS/123-QED}

\title{Ferromagnetic nano-structures in Valenta model}

\author{I. Zasada}
\email{IZASADA@wfis.uni.lodz.pl}
\author{B. Busiakiewicz}
 \affiliation{Solid State Physics
Department, University of Lodz, ul. Pomorska 149/153, Lodz,
Poland}

\date{\today}

\begin{abstract}
The ferromagnetic nano-structures are recently of great interest
for modern investigations. A comparison of the experimental data
and theoretical results shows that the use of the standard
molecular field approximation is insufficient for the description
of nano-structure properties. Therefore, we use the effective
field approach in order to show the usefulness of the Valenta
model generalized in this way. The agreement between experiment
and theory is then excellent.

The magnetization profiles and the calculated Curie temperatures
are presented for the systems consisting of Ni and Co layers with
different configuration of the surfaces and interfaces including
terraces and wires. We have shown that the position in the system
as well as the kind of neighbouring layers and their mutual
interactions can determine the shape of magnetization profiles.
The use of the Valenta model allows us to present all dependences
in the layer resolved mode.

\end{abstract}

\keywords{thin films, nano-structures, magnetic phase transition}
\maketitle

\section{Introduction}

A recent development of nanotechnologies needs still more actual
and more effective methods for the description of nano-structures.
In this case the ultra-thin films are very convenient systems in
order to describe their properties as a starting point towards the
nano-objects.

In the above circumstances the description of magnetic thin films
is of fundamental significance from the theoretical point of view
because of their role in the interpretation for the thermodynamics
applied to inhomogeneous media and for the quantum mechanics of
objects with restricted dimensions.

The problem of the thin films theory is rather old, but the model
introduced by Valenta fifty years ago \cite{b1,b2} seems to be
still an excellent and effective approach for the characterization
of the magnetization properties and the phase transition
temperature. The model formally equivalent to molecular field
approximation is based on broader physical background and
represents an original interpretation from the physical point of
view. The model in its extended form allows us to discuss not only
the magnetic properties but also the lattice thermodynamic
\cite{b3}, order-disorder \cite{b4}, electronic phenomena
\cite{b5} as well as the phase transitions in the context of the
stability and surface melting \cite{b6}.

Recently, a new attempt at the study of different systems and
phenomena within the extended Valenta model has been made.

First of all, different properties of $AB_{3}$ binary alloy thin
films have been discussed. The long-range order parameter and the
concentration distribution, mutually dependent and dependent on
the film thickness and the boundary conditions at the surfaces
have been analyzed \cite{b7,b8}. An interpretation of the
disordering kinetics has been proposed in connection with the
behaviour of disorder which appears as layer by layer process
starting form the surface plane or from the sample inside in
dependence on the boundary conditions. An interdependence between
the surface melting and the surface disordering observed in binary
alloy thin films has been considered in the context of their
mutual relations \cite{b9,b10}. The presented approach allowed to
extend the phase transitions diagrams to the case when the
crystallinity parameter behaviour influences the local
concentration profiles and the lattice order parameter describing
the alloy structure. In particular, the surface-induced disorder
was described when the crystal structure is preserved and, in
contrast, when the surface melting is expected for partially
disordered samples. The behaviour of the magnetization in the
binary alloy thin films with respect to their lattice disorder and
the relative concentration of alloy components has been also
discussed \cite{b11}. An interesting fact is observed when the
magnetic order appears for the lattice disorder, i.e. for higher
temperatures while there is no magnetization for the
lattice-ordered state. The order-disorder phenomena including the
distribution of chemical component has been considered as well in
terms of the short-range order in binary alloy thin films whose
surfaces play an essential role \cite{b12}. Among others, two
effects seem to be of particularly great interest, namely: the
crossover of the site occupancy in the surface layer and the shift
of atoms in the surface plane between two kinds of lattice sites.
Such effects influence the diffuse low-energy electron
diffraction, surface melting or spin-wave resonance conditions.

Next, different levels of extensions for the original Valenta
model have been proposed in connection with its applicability for
magnetic thin films description. The most important one is
connected with the improvement of the entropy construction in the
self consistent way when the correlations are taken into account
\cite{b13}. Another kind of extension is connected with
introduction of the reaction field instead of the standard
molecular field. The procedure for the reaction field approach
(RFA) consists in adding to the molecular field a correlation
dependent term determined by means of the fluctuation-dissipation
theorem. The part of the effective field arising from the reaction
field does not favor one orientation over another while the
molecular field is directed along the spontaneous magnetization
axis. The Valenta model generalized in such a way has been applied
for the description of ferromagnetic films separated by the
nonmagnetic spacer forming the multilayer systems \cite{b14}.

In the present paper, we show that the Valenta model can be also
successfully applied to different ferromagnetic nano-structures.
We start with ultra-thin multilayer systems, next we consider the
systems with terraces and at the end we discuss the wires
properties in connection with different environment conditions.

\section{The background of the Valenta model}

The model introduced by Valenta in its original formulation is
formally equivalent to the molecular field approximation, and for
this reason, it is at present treated as an insufficient approach
for the precise description of thin films properties. However,
taking into considerations the original assumptions which are
based on the deep physical background, we can modify the model in
various aspects with respect to the purposes of the expected
interpretations.

    The Valenta model \cite{b1,b2}consists in two assumptions:

- the discretization of sample geometry which reflects the
crystallographic lattice structure and some surface irregularities
connected with the nearest neighbours for each of the surfaces.
The discrete structure of variables, in particular, the distances
between the neighbouring planes in the film thickness direction,
leads to the formulation of the equations describing the film in
the form of the difference equations with properly chosen boundary
conditions.

- the thermodynamics modified for inhomogeneous systems. It turns
out that a sample with thin film geometry is such an inhomogeneous
system from the thermodynamic point of view, so it should be
considered in terms of thermodynamics modified for small clusters
\cite{b15} or nano-structures \cite{b16}.

A film can be treated as a sample cut in some crystallographic
orientation with respect to the surface of the crystal with a
given crystallographic structure characterized by the spectrum of
the nearest neighbouring atoms. In this case the atoms situated at
the surfaces have their neighbourhood which is different with
respect to the considered site vicinity inside a sample. The
geometric situation corresponds then to the different conditions
in which the atoms at the surface and the atoms inside a sample
are embedded. However, the film can be divided into monoatomic
layers parallel to the surface planes from which a layer in two
dimensions can be treated as thermodynamically homogenous
subsystem. From the thermodynamic point of view a film is then
interpreted as a composition of homogenous subsystems in the sense
of N\'{e}el sublattices \cite{b17}. The N\'{e}el idea consists
here in the division of the system into several groups,
sublattices, which form the homogenous subsystems gathering the
physically equivalent particles, i.e., equal particles accruing in
the physically identical conditions. The N\'{e}el idea applied by
Valenta to thin films can be interpreted that the inhomogeneous
system can be divided into homogenous subsystems which are
identical with monoatomic layers parallel to the surfaces,
however, embedded into different neighbourhood.

The division of a thin film into homogeneous sublattices which are
monoatomic layers parallel to the surface implies in terms of
quantum mechanics the assumption \cite{b1,b2} that the total wave
function of the electrons of a thin film practically differs very
little from the wave function of the state in which the components
of magnetic moments in the monoatomic layers have well defined
values. In these conditions, however, the model corresponds to the
situation when the hamiltonians of the subsystems do not commute
with the hamiltonian of the system. Although the commutative rules
can only be approached, they are satisfied in a very simple case
when the interactions between the sublattices are determined by
means of the effective field.

Thus, at high temperature, the quantum mechanical construction
influences the thermodynamic interpretation. The hamiltonian of
each Néel sublattice corresponds to the integral of motion for
each layer. This means that the total wave function of the state
is the product of wave functions with respect to an individual
sublattice. From the thermodynamic point of view this fact is
equivalent to the factorization of the partition functions. The
total partition function is factorized with respect to the
partition functions of individual monoatomic layers. The
statistical operator of a film is then a product of the
statistical operators of layers due to the additive character of
the effective hamiltonians. As a consequence, the entropy is a sum
of terms describing homogeneous contributions of monoatomic layer
entropies independent of one another in calculations. At low
temperatures the construction of a Heisenberg type quantum
mechanical theory leads to the solution which can be related to
the spin waves propagation or, in the magnon representation, to
the quasi free particles when they are embedded in the heat bath
of harmonic oscillators. This level of approximation corresponds
to the case when the transversal correlations between spins are
neglected. Then, the hamiltonian is reduced to the Ising type
hamiltonian whose longitudinal correlations reduce to the MFA
results. In the thin films geometry the direction perpendicular to
the surface plane is distinguished in a natural way by breaking
the translational symmetry whose perturbation gives the conditions
in which the size effects can be observed. One of the fundamental
characteristics of the considered model is a discretization of
variables, at least, in the film thickness direction. The
geometrical properties of a film lead to their description by
means of the variational procedure which should be considered in
the discrete space. The variational equations are then of
difference, not differential, forms.

The thermodynamic approach is, in general, based on the free
energy functional construction:
\begin{equation}
F=U-TS \label{e1}
\end{equation}
which can be obtained by means of the internal energy $U$  and the
entropy  $S$ calculations. The internal energy $U$ is defined by
the mean value of the Hamiltonian describing the considered system
while the entropy in the pair representation is then given in the
standard form \cite{b13,b18}:

\begin{equation}\label{e2}
S= \sum_{\nu} \sigma_{\nu}+\sum_{<\nu \nu'>}( \sigma_{\nu
\nu'}-\sigma_{\nu} -\sigma_{\nu'})
\end{equation}
where
\begin{eqnarray}\label{e3}
\sigma _{\nu}&=& k_{B} \Bigg [ \Big ( \frac {1}{2} +m_{\nu} \Big )
\ln \Big ( \frac {1}{2} +m_{\nu} \Big ){} \nonumber\\
&&{}+\Big ( \frac {1}{2} -m_{\nu} \Big ) \ln \Big ( \frac {1}{2}
-m_{\nu} \Big ) \Bigg ]
\end{eqnarray}
We can see from (\ref{e2}) that the entropy has then two
contributions. One of them corresponds to the single-site entropy:
\begin{equation}\label{e4}
S_{1}= N^{2} \sum_{\nu} [1-z(\nu)]\sigma_{\nu}
\end{equation}
for $z(\nu)$  denoting the number of nearest neighbours when the
central site lies in the plane $\nu$, namely $z(\nu)=\sum_{\nu'}
z(\nu,\nu')$  where $z(\nu,\nu')$ stands for the number of nearest
neighbours in the plane $\nu'$  when the central site is in the
plane $\nu$. $N$ stands for the number of spins in the linear
dimension of the plane ($nN^{2}$ denotes the number of the spins
in the system consisting of $n$ layers).

The contribution of pair entropy can be written as:
\begin{equation}\label{e5}
S_{2}= S_{2}^{0} + S_{2}^{1}
\end{equation}
where, the first term contains the contribution introduced in the
planes  $\nu$, i.e. in the homogeneous subsystems, namely:
\begin{equation}\label{e6}
S_{2}^{0}= \frac{1}{2}N^{2} \sum_{\nu} z(\nu,\nu)\sigma_{\nu\nu}
\end{equation}
and, the second one contains the contribution introduced by the
interactions between two monolayers, i.e. the interactions between
subsystems embedded into inhomogeneous bath, namely:
\begin{equation}\label{e7}
S_{2}^{1}= \frac{1}{2}N^{2} \sum_{\nu\nu' \ne \nu}
z(\nu,\nu')\sigma_{\nu\nu' \ne \nu}
\end{equation}
The entropy in each monoatomic layer, following the calculations
for homogenous system, is of the form
\begin{eqnarray}\label{e8}
\sigma _{\nu\nu}& =& k_{B} \bigg[\Big(\frac{1}{4} + m_{\nu} +
c_{\nu}\Big)\ln \Big(\frac{1}{4} + m_{\nu} + c_{\nu}\Big) {}\nonumber \\
& & {}+ 2\Big(\frac{1}{4} -  c_{\nu}\Big) \ln \Big(\frac{1}{4} -
c_{\nu}\Big){}\\
& & {} +\Big(\frac{1}{4} - m_{\nu} + c_{\nu}\Big)\ln
\Big(\frac{1}{4} - m_{\nu} + c_{\nu}\Big)\bigg]\nonumber
\end{eqnarray}
while, the entropy contribution given by the interlayer
interactions leads to:
\begin{eqnarray}\label{e9}
 \sigma _{\nu\nu' \ne \nu} & = & k_{B}
\bigg[\Big(\frac{1}{4} +\frac{1}{2} (m_{\nu} +m_{\nu'}) + c_{\nu\nu'} \Big) \times{}\nonumber\\
& & {} \times \ln \Big(\frac{1}{4} + \frac{1}{2} (m_{\nu}
+ m_{\nu'})+ c_{\nu\nu'}\Big)+ {}\nonumber \\
& & {}+ \Big(\frac{1}{4} +
\frac{1}{2} (m_{\nu} - m_{\nu'}) - c_{\nu\nu'} \Big) \times {}\nonumber\\
& & {} \times \ln \Big(\frac{1}{4} + \frac{1}{2} (m_{\nu}
- m_{\nu'})- c_{\nu\nu'}\Big)+ {}\nonumber \\
& & {} + \Big(\frac{1}{4} -
\frac{1}{2} (m_{\nu} - m_{\nu'}) - c_{\nu\nu'} \Big) \times {}\\
& & {} \times \ln \Big(\frac{1}{4} - \frac{1}{2} (m_{\nu}
- m_{\nu'})- c_{\nu\nu'}\Big)+ {}\nonumber \\
& & {}+ \Big(\frac{1}{4} -
\frac{1}{2} (m_{\nu} + m_{\nu'}) + c_{\nu\nu'} \Big) \times {}\nonumber\\
& & {} \times \ln \Big(\frac{1}{4} - \frac{1}{2} (m_{\nu} +
m_{\nu'})+ c_{\nu\nu'}\Big) {}\nonumber
\end{eqnarray}

When we consider the quantities $\sigma_{\nu}$, $\sigma_{\nu\nu}$,
and  $\sigma_{\nu\nu' \ne \nu}$, the factorization of the
statistical operator is evident for  $\sigma_{\nu}$ and
$\sigma_{\nu \nu}$while the factorization for $\sigma_{\nu\nu' \ne
\nu}$  is not obvious because of the interactions between the
sites localized perpendicularly to the monoatomic layers. It is
worth-while to notice here that the factorization of the
statistical operator is equivalent to the additive character of
the entropy. The simplest illustrative example confirming the
above statement can be given by the assumption that $c_{\nu\nu'
\ne \nu}=m_{\nu}m_{\nu'}$  which leads to the factorized term (9)
in a self-consistent way. Taking into account the above assumption
we can consider the pair entropy (\ref{e6}) only in the planes,
while the entropy contribution due to the nearest neighbouring
layers interactions is reduced to the single-site entropy term.

The equilibrium values of the layer dependent magnetic order
parameters $m_{\nu}$  are obtained by the variational procedure
minimizing the free energy (\ref{e1}) with respect to the
parameters $m_{\nu}$. The obtained set of equations can be solved
numerically leading to the layer and temperature dependent
magnetization; i.e. its profile across a film as well as the Curie
temperature evaluation.

\section{Nano-structures in the Valenta model}
Originally introduced for magnetic films, the Valenta model
outlined above is now generalized for different systems in which
their inhomogeneity is taken into account. The Valenta model
assumes the infinite dimensionality in the surface plane. In this
context we can speak about the nano-dimension only in the
direction perpendicular to the surface and multilayers consisted
of several monoatomic planes can be considered as nano-objects
with restricted dimension in one direction. We can, however,
introduce one or two edges in the surface plane forming terraces,
wires or atomic chains, nano-structures homogeneous only in one
direction which is in perfect accordance with the original Neel
idea of sublattices. We consider now, the example calculations for
the systems mentioned above.

\subsection{Ferromagnetic multilayers}

The magnetic properties of multilayer systems consisting of two
different kinds of magnetic films separated by nonmagnetic spacer
are widely studied from the experimental as well as theoretical
point of view \cite{b19,b20,b21,b22,b23}. The ferromagnetism in
such systems is usually discussed in terms of the temperature
dependent long-range order parameter i.e. the spontaneous
magnetization. From such dependence the behaviour of the Curie
temperature can be derived and analyzed.

It has been theoretically calculated and experimentally shown
\cite{b14,b19} that for two ferromagnetic Ni and Co films coupled
by the indirect exchange interaction  $J_{inter}$ (via the Cu film
- Co/Cu/Ni/Cu(100) system) one obtains two different ordering
temperatures and we can observe two susceptibility signals
\cite{b22} one in $T_{C,Co}$  and a weaker one in $T_{C,Ni}^{*}$.
Then, we treat the  $T_{C,Ni}^{*}$ as a quasi-critical phase
transition temperature of the system. It has been also shown
\cite{b24} that such system can exhibits an inverse behaviour. By
selecting the appropriate thicknesses of Co, Cu and Ni films, we
can observe the lower ordering temperature of Co than the one of
Ni.

We present here the numerical calculations for these two different
experimental situations performed in the frame of RFA \cite{b14}.
In this case the internal energy $U$  is given in the following
form:
\begin{eqnarray}\label{e10}
U & = & -N^{2}z_{0} \sum _{\nu}J_{\nu \nu} m_{\nu}m_{\nu}  {}\nonumber \\
& & {}-z_{1} \sum _{\nu'\in \nu} \frac{J_{\nu \nu'}}{2} m_{\nu}m_{\nu \pm 1}  {}\nonumber \\
& & {}- \sum _{\nu} (K_{\nu} - \lambda) m_{\nu} m_{\nu } {}\nonumber \\
& &{} -\gamma \sum _{\nu }H_{\nu}^{z} m_{\nu}
\end{eqnarray}
and the entropy  $S$ is expressed by equation (\ref{e2}) reduced
to its single site representation.  $z_{0}$ and  $z_{1}$ are the
numbers of nearest neighbours of a given atom in the same
monoatomic layer and in the neighbouring layers and for fcc(100)
structure they are,  $z_{0}=4$ $z_{1}=4$, respectively; $J_{\nu
\nu'}$ represents the exchange integral responsible for the
interaction between a given spin and its nearest neighbours in the
same magnetic layer ($\nu = \nu'$) or in the neighbouring layers
($\nu' = \nu \pm 1$), the index numbers the monoatomic layers of
the thin film ($\nu = 1, ..., n$). It is convenient to denote
$J_{\nu \nu'}$ as $J_{Ni}$ and $J_{Co}$ for interior Ni and Co
layers while for nickel layers which are in direct connection with
copper we put an interface exchange coupling $J_{Ni}^{interface}$.
The Co/Cu interface is not differentiated because it is
approximately compensated by the enhancement of the magnetic
moment in the topmost layer facing the vacuum \cite{b25} and we
take the same values of the exchange integral for all monoatomic
layers forming the Co film. The interaction between the Co and Ni
films in the Co/Cu/Ni/Cu(100) system takes place via the
interlayer exchange coupling  $J_{inter}$, which depends on the
nonmagnetic spacer thickness \cite{b14}. The anisotropy constant
$K_{Ni}$ including volume and the surface anisotropy term, can be
distinguished as we did in the case of the exchange coupling,
namely in each magnetic film we have in-plane anisotropy $K_{Ni}$
and $K_{Co}$ while at the interface Ni/Cu we have
$K_{Ni}^{interface}$. Additionally, in order to take into account
the boundary conditions connected with the surfaces state we
consider also a perpendicular anisotropy $\kappa_{S}$  which is
included in the parameter $\lambda$ calculations \cite{b14}. The
parameter $\lambda$ appearing in equation (\ref{e4}) is the
correlation parameter characteristic for RFA which is independent
of $(\nu)$ due to symmetry conditions \cite{b26} and it is assumed
to be homogeneous in the sample. This parameter is entirely
determined by the crystallographic structure of the considered
sample and it is not the additional parameter of the theory. For
its calculation one needs only the values of exchange integrals
and anisotropy constants mentioned above and of course the
crystallographic data connected with the nearest neighbours
distribution.

According to the above discussion and on the bases of the
experimental findings \cite{b19} we have chosen the following
values for the exchange integrals and anisotropy constants:
$J_{Ni} = 1.7 \cdot 10^{-21} $J, $J_{Ni}^{interface} = 0.5
J_{Ni}$, $J_{Co} = 3 \cdot 10^{-21}$J, $J_{inter} = -1.82 \cdot
10^{-23} $J, $K_{Ni} = 0.001 \cdot  J_{Ni}$, $K_{Ni}^{interface} =
0.001 \cdot J_{Ni}^{interface}$, $K_{Co} = 0.001 \cdot  J_{Co}$
and $\kappa_{S} = -1$.

\begin{figure}
\includegraphics[scale=0.7]{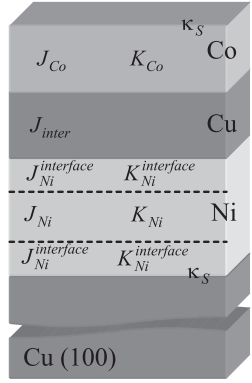}
\caption{\label{fig1}The trilayer system Co/Cu/Ni/Cu(100) with the
exchange couplings and anisotropies used in the theoretical
model.}
\end{figure}

\begin{figure}
\includegraphics[scale=0.7]{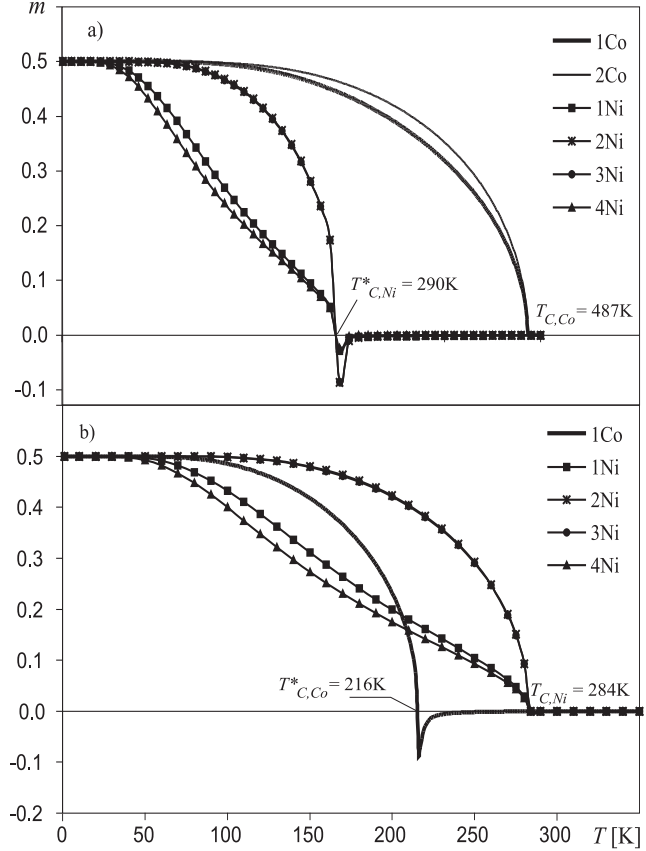}
\caption{\label{fig2}The layer-dependent magnetization as a
function of temperature determined for 2ML Co/2ML Cu/4ML
Ni/Cu(100) system (a) and  1ML Co/2ML Cu/4ML Ni/Cu(100) system
(b).}
\end{figure}

The schematic view of the discussed multilayer system is presented
in Fig. \ref{fig1}. First, we consider the 2ML Co/2ML Cu/4ML
Ni/Cu(100). The temperature dependent layer resolved magnetization
of this system is shown in Fig. \ref{fig2}a and we can see that
for the whole range of temperatures the magnetization of Ni film
is much smaller than for Co one, which is caused by the difference
in the values of magnetic moments reported for these two materials
\cite{b19}. If we now select a Co thickness of only 1ML, the
ordering temperature of Co is below that of Ni. In Fig.
\ref{fig2}b we present the layer-resolved temperature dependent
magnetization for the 1ML Co/2ML Cu/4ML Ni/Cu(100) system. It is
clearly seen that in this case we observe the quasi-critical
temperature for Co layer $T_{C,Co}^{*}$  and the usual phase
transition in the Ni film transition temperature $T_{C,Ni}^{*}$.
We have, however, to underline that one monoatomic layer deposited
on the Cu(100) substrate is determined by different physical
conditions than the thin Co film containing 2ML on Cu(100). In the
case of 1ML Co we cannot speak any more about magnetic moment
compensation at surface and interface and we have to take into
account the decrease
of the magnetic moments at the interface (~32
exchange integral from  $J_{Co} = 3 \cdot 10^{-21}$J to $J_{Co} =
1.8 \cdot 10^{-21}$J \cite{b25}.

\begin{figure}
\includegraphics[scale=0.6]{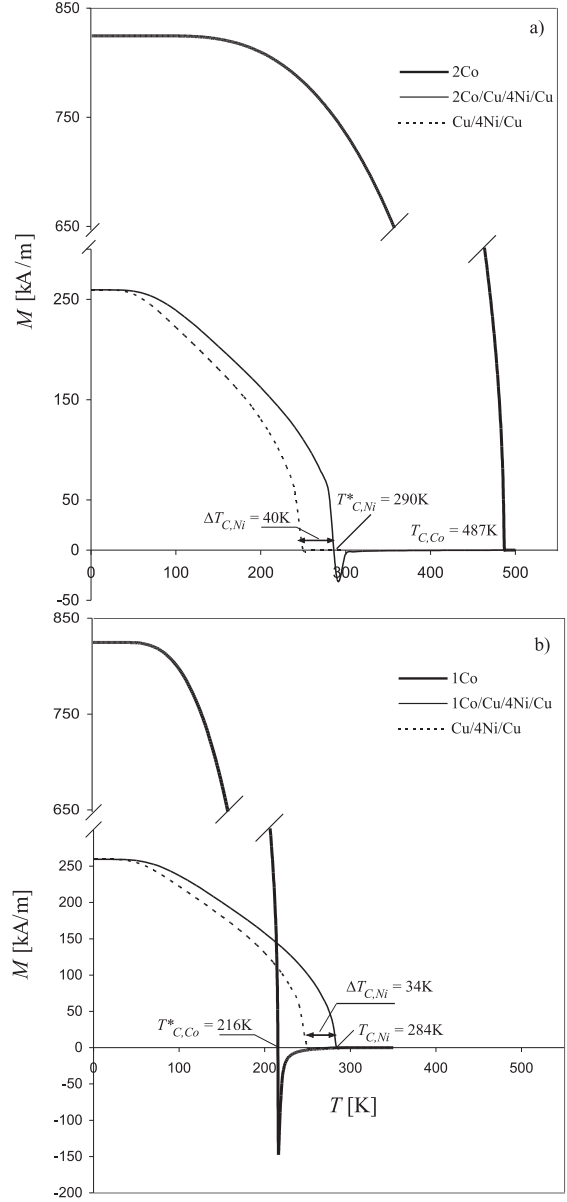}
\caption{\label{fig3}Average Ni magnetization of a bilayer Ni film
(dashed line (a) and (b)) and of Ni film (thin solid line) coupled
to Co film in 2ML Co/2ML Cu/4ML Ni/Cu(100) system (a) and  1ML
Co/2ML Cu/4ML Ni/Cu(100) system (b). The thick solid line ((a) and
(b)) refers to the Co magnetization.}
\end{figure}

Fig. \ref{fig3} shows the magnetization curves obtained as an
average over all Ni and Co layers forming the Ni and Co films in
the trilayer systems 2ML Co/2ML Cu/4ML Ni/Cu(100) and 1ML Co/2ML
Cu/4ML Ni/Cu(100) in Fig. \ref{fig3}a and Fig. \ref{fig3}b,
respectively. In both graphs we also show the average
magnetization curve for the Ni film composed of 4ML and embedded
in the bilayer system 2ML Cu/4ML Ni/Cu(100). In this case we have
different, with respect to the trilayer system, boundary
conditions characterizing the whole system surfaces state. In
order to take it into account we have put a perpendicular
anisotropy of different value, namely $\kappa_{S} = -0.1$. In both
trilayer systems we observe the tails of remanent magnetization of
Ni film for the first one and of Co film for the second one. The
Ni and Co remanent magnetization changes sign at the
quasi-critical temperature $T_{C,Ni}^{*}$ and $T_{C,Co}^{*}$,
respectively, due to the antiferromagnetic coupling caused by the
copper spacer of 2 ML thickness \cite{b14}.

The Co film in 1ML Co/2ML Cu/4ML Ni/Cu(100) system has lower
ordering temperature than the Ni and its magnetization is
vanishing in the presence of ferromagnetic Ni film. Due to the
fact that the ground state moment of Co is larger than the Ni one,
the element-specific $M(T)$  curves for Co and Ni cross each other
close to $T_{C,Co}^{*}$ (Fig. \ref{fig3}b). Such crossover can be
interpreted as connected with the Ni magnetization rotation from
parallel (at low temperatures) to perpendicular (at
$T_{C,Co}^{*}$) and antiparallel (up to $T_{C,Ni}$) to the Co one.

In both considered cases the magnetization of Ni film from the
trilayer system is higher then its magnetization in bilayer system
and the Curie temperature is shifted toward the higher values by
$\delta T_{C,Ni}=40K$ in trilayer with 2ML Co and $\delta
T_{C,Ni}=34K$ in trilayer with 1ML Co. The difference between
these two shifts is due to interlayer exchange coupling
$J_{inter}$ which enhances the Ni magnetization in the case of 2ML
Cu/4ML Ni/Cu(100).

The magnetization behaviour depends not only on the composition
but also on the thickness of a system. It is clearly seen in Fig.
\ref{fig2} and Fig. \ref{fig3}. Thinner system has lower phase
transition temperature. In the case when the system is composed of
two films with different transition temperatures we can expect
that if we choose the relative thicknesses of Co and Ni films in a
way that they have the same critical temperature the trilayer
system will exhibit only one common Curie temperature (without
quasi-critical one). It is indeed observed experimentally
\cite{b24}.

\subsection{Ferromagnetic terraces}
\begin{figure}
\includegraphics[scale=0.7]{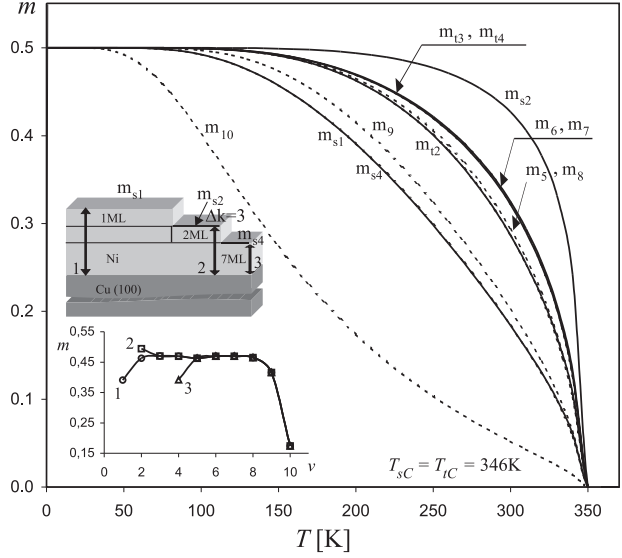}
\caption{\label{fig4}The layer-dependent magnetization as a
function of temperature determined for the system with two
terraces supported by 7ML Ni in configuration shown in the inset.
The second inset shows the magnetization profiles across the first
terrace region (curve 1), second terrace region (curve 2) and the
rest of a sample (curve 3).}
\end{figure}

\begin{figure}
\includegraphics[scale=0.7]{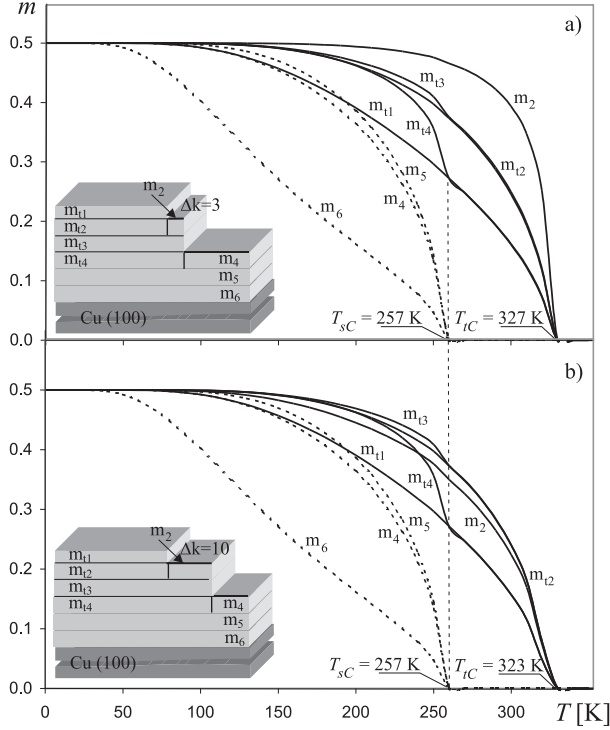}
\caption{\label{fig5}The layer-dependent magnetization as a
function of temperature determined for the system with two
terraces supported by 3ML Ni in configurations shown in the
insets.}
\end{figure}

Let us now consider a simple layered structure with two terraces
shown in the insets of Fig. \ref{fig4} and Fig. \ref{fig5}. The
first terrace is semi-infinite in plane and is composed of 1
monoatomic layer while the second one is large of   and is
composed of 2 monoatomic layers. The surface of the second terrace
is then restricted in two dimensions and has two different
boundary conditions at the edges. One of its edges is interacting
with the first terrace while another one is exposed to vacuum.
Both terraces are supported by several infinite layers put on-top
of non-magnetic material forming Ni/Cu(100) system. In this case
the system is divided in several different homogeneous subsystems
and it has to be underlined that $2^{nd}$ and $4^{th}$ monoatomic
layers are divided into two different subsystems each. The first
group of subsystems, from $5^{th}$ to $10^{th}$ monoatomic layers,
consists of infinite planes parallel to the surface. The second
group, from $2^{nd}$ to $4^{th}$ monoatomic layers, consists of
semi-infinite planes parallel to the surface. The third group,
s1st and $s^{4th}$ monoatomic layers, consists of semi-infinite
surface planes, each of them having, however, different boundary
conditions at their edges. And finally the last subsystem,
$s2^{nd}$ monoatomic layer, forms the plane infinite in only one
dimension having two different boundary conditions for each of its
edges. In Fig. \ref{fig4} we show the temperature dependent layer
resolved magnetization for two Ni terraces ($\Delta k=3$)
supported by 7ML Ni film. Such system exhibits one phase
transition temperature $T_{sC} = T_{tC} = 346$K and different
magnetization distributions in the direction perpendicular to each
of three surface planes (inset in Fig. \ref{fig4}).

Much more interesting situation appears, however, when the same
two terraces are supported by only 3 ML of Ni/Cu(100). We observe
then two transition temperatures lower one for the Ni support and
the second one, higher, for terraced Ni structure. In Fig.
\ref{fig5} we present the results obtained for two widths of the
second terrace, $\Delta k=3$ and  $\Delta k=10$, respectively in
Fig. \ref{fig5}a and Fig. \ref{fig5}b. In both cases the Curie
temperature for the support has the same value $T_{sC} = 257$K
while the transition temperature for the terraced part of the
sample depends on the second terrace width being higher for
$\Delta k=3$. Moreover, at the support Curie temperature we can
notice the jump of the magnetization value for $m_{t3}$ $m_{t4}$.
Above this temperature the magnetization curves $m_{t3}$ and
$m_{t4}$ become identical with $m_{t2}$ and $m_{t1}$ magnetization
curves, respectively (Fig. \ref{fig5}). We can also notice that
the position of magnetization curve $m_{2}$ with respect to the
rest of the curves depends on $\Delta k$. For smaller value this
magnetization exhibits the dominant role having the highest values
in the whole range of temperatures.

\subsection{Ferromagnetic wires}
\begin{figure}
\includegraphics[scale=0.8]{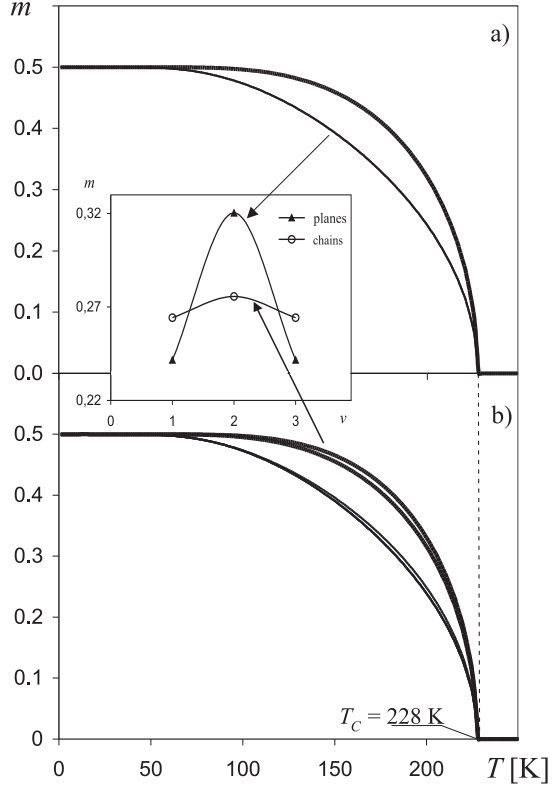}
\caption{\label{fig6}The temperature dependent magnetization
determined for two types of   wires: (a) constructed with
monoatomic planes and (b) constructed with monoatomic chains. The
inset shows the magnetization profiles for both systems (see
description in the text).}
\end{figure}

\begin{figure}
\includegraphics[scale=0.8]{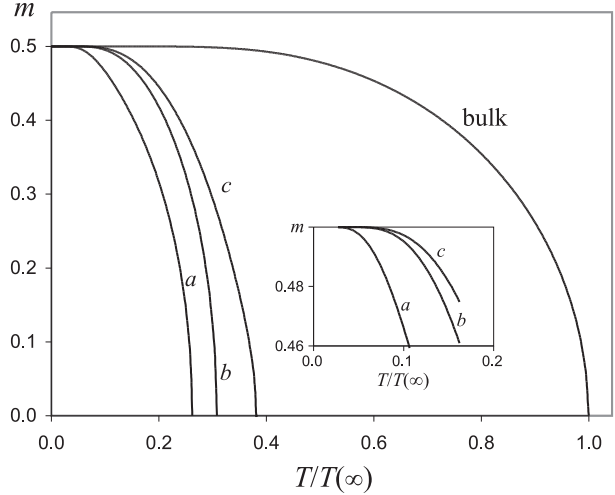}
\caption{\label{fig7}The temperature dependent magnetization
determined for three nanostructures: free-standing wire (b); the
same wire deposited on 1Ml of Ni (c); the same wire deposited on
1Ml of Cu (a). The inset shows the low-temperature magnetization
behavior for all Ni nanowires}
\end{figure}

In the frame of the Valenta model a wire can be constructed in two
different ways. First of all, it can be composed with the
monoatomic planes infinite in one direction with symmetric
boundary conditions at both in-plane edges. The size of such wire
is then determined by the number of planes and their width. The
second way consists in dividing the above described wire in
monoatomic chains; it means the homogeneous subsystems infinite in
one direction. Fig. \ref{fig6} shows the temperature dependent
magnetization curves calculated for both wire types of $3 \times
3$ size. The Curie temperature is common but the magnetization
distribution differs which can be clearly seen in the inset where
the average magnetization for three in-plane monoatomic chains is
compared with the magnetization of monoatomic planes forming the
wire of the first type. We can then discuss which construction is
proper. It seems that the monoatomic chains give a better
resolution in magnetization calculations and its distribution is
more reliable but of course the experimental results are decisive.
In Fig. \ref{fig7} we present the temperature dependent average
magnetization calculated for three nanostructurs: free-standing
wire composed with 9 monoatomic chains in $3 \times 3$
arrangement; the same wire deposited on 1Ml of Ni; the same wire
deposited on 1Ml of Cu. The temperature dependent magnetization of
Ni bulk is shown for comparison and the inset shows the
low-temperature magnetization behavior for Ni nanowires examined
in three different environments. The dependence of the
magnetization of Ni nanowire is qualitatively equivalent to the
magnetic properties of Ni nanowires electrodeposited into
self-assembled alumina arrays reported in paper \cite{b27}.

\section{\label{sec:level4}  Concluding remarks}
The interest of various investigations and applications of
ferromagnetic nano-objects in nanotechnology is a factor
stimulating the development of their theoretical description in
which the methodology plays an essential role. In the
methodological context, the model of magnetic thin films
introduced by Valenta \cite{b1,b2} seems to be one of the most
important methods which respect the thermodynamics of
inhomogeneous systems formulated in the Néel representation.

It is worth while to mention that the Valenta model is of the
Heisenberg type treatment. By this fact, it allows us to consider
the spin wave description in particular, the spin wave resonance
as well as the magnetization at low temperatures. In order to
discuss the high temperature behaviour of magnetization, the
effective molecular field is applied. Usually this approach
reduces the hamiltonian to its Ising forms. However, the use of
the reaction field approximation (RFA) takes also the transversal
components of the spin operators into account. In this context the
considerations presented in \cite{b14} and applied in the present
paper bring the background for the nano-objects treatment in terms
of RFA method from the point of view of their entropy formulation
which should be then taken at least in the pair representation.
The RFA approach is very sensitive for the description of systems
with restricted dimensions due to the anisotropic character of the
basic hamiltonian. The isotropic case corresponds then to the
phase transition temperature equal to zero in agreement with the
Mermin-Wagner theorem predictions.

Similar evaluations can be discussed in the case of correlations
introduced by the constant coupling approach \cite{b28}, the
Bethe-Peierls-Weiss method \cite{b29}, as well as the Green
function technique \cite{b30} which is common for low and high
temperature regions, leading to the magnetization profiles found
in a self-consistent way \cite{b31}. We have shown \cite{b14} that
the equivalence between the results of the Valenta model and the
Green functions approach in the random phase approximation for the
Ising model is evident as their property of general character. The
extension of the Valenta model by the introduction of the reaction
field correction corresponds to the situation in which the spin
correlations are taken into account. Although this idea is common
for the considerations within the Valenta model and the Green
functions approach, as far as the physical interpretation is
discussed, the results are now not identical due to the different
description of the magnetization profiles.

Originally, the Valenta model is constructed for the homogeneous
angular distribution of magnetic moments oriented in agreement
with the quantization axis. We can observe, however, two
orientations at least, parallel and perpendicular to the surface.
Taking into account this fact, we can extend the Valenta model to
the discussion concerning the angular distribution of
magnetization in terms of the variational procedure derived in the
discrete space. We obtain then not only two orientations of
homogeneous distribution but also their inhomogeneous distribution
among the sublattices \cite{b32}.

An important generalization of the Valenta model is still expected
in the characterization of the surfaces. The surface roughness and
surface alloying are among the most interesting phenomena to be
considered within the Valenta model. The surface roughness is a
determinant of the surface texture and it appears in every scale.
The spontaneous formation of defects in atomic scale on the solid
surface is expected for purely entropic reasons. The study of
these defects is fundamental for the surface science as well as
for understanding of crystal growth, catalysis and diffusion
mechanisms. The surface roughness description based on the local
statistical identities leads to physically interesting properties
of the surface in the context of surface phase transitions. Thus
the topical use of the Valenta model is also connected with the
interpretation of the STM image.

At the end, it is worth while to notice that the thermodynamic
approach to the description of nano-structures within the Valenta
model gives the information about the temperature dependence of
divers order parameters describing the inhomogeneous system as
well as allows to distinguish the properties of any particular
homogeneous subsystem which can be defined on the basis of
N\'{e}el concept of sublattices. Moreover, the Valenta model
construction shows the predictions for the methodological
formulation of the thermodynamic of inhomogeneous small systems.

The particular results considered in the present paper provide an
illustrative example whose generalization seems to us to have a
general meaning for the discussion of the methods and their
comparison at different levels of accuracy.

\begin{acknowledgements}
The authors are grateful to Professor Leszek Wojtczak, a close
friend and coworker of Professor Luboš Valenta, for his
stimulating discussions during the study on Valenta model.
\end{acknowledgements}


\begin{thebibliography}{99}
\bibitem{b1}L. Valenta, Czech. J. Phys. \textbf{7},  127 (1957).
\bibitem{b2}L. Valenta, Czech. J. Phys. \textbf{7},  136 (1957).
\bibitem{b3}L. Wojtczak, S. Zajac, Bull. Acad. Polon. Sci. \textbf{16},  527 (1966).
\bibitem{b4}L. Valenta, A. Sukiennicki, Phys. Stat. Sol. \textbf{17},  903 (1966).
\bibitem{b5}L. Valenta, L. Wojtczak, Czech. J. Phys. B\textbf{30},  1025 (1980).
\bibitem{b6}L. Valenta, Czech. J. Phys. \textbf{46},  607 (1996).
\bibitem{b7}F. L. Castillo Alvarado, A. Sukiennicki, L. Wojtczak, I. Zasada, Physica B \textbf{344},  477 (2004).
\bibitem{b8}I. Zasada, L. Wojtczak, S. Mróz, J. Alloys Compd. (2008), doi:10.1016/j.jallcom. 2008.02.015.
\bibitem{b9}L. Wojtczak, I. Zasada, A. Sukiennicki, F. L. Castillo Alvarado, Phys. Rev. B \textbf{70},  195416 (2004).
\bibitem{b10}L. Wojtczak, I. Zasada, T. Rychtelska, Surf. Sci. \textbf{600},  851 (2004).
\bibitem{b11}A. Sukiennicki, L. Wojtczak, I. Zasada, F. L. Castillo Alvarado, JMMM \textbf{288},  137 (2005).
\bibitem{b12}I. Zasada, A. Sukiennicki, L. Wojtczak, F. L. Castillo Alvarado, Phys. Rev. B \textbf{74},  205402 (2006).
\bibitem{b13}I. Zasada, B. Busiakiewicz, L. Wojtczak, JMMM \textbf{312},  58 (2007).
\bibitem{b14}B. Busiakiewicz, I. Zasada, L. Wojtczak, J. Phys.: Condens. Matter \textbf{20},  095217 (2008).
\bibitem{b15}T. L. Hill, J. Chem. Phys. \textbf{36},  3182 (1962).
\bibitem{b16}H. Puszkarski, A. R. Ferchmin, Acta Physicea Superficierum \textbf{4},  55 (2001).
\bibitem{b17}L. Néel, Ann. de Phys. \textbf{3},  137 (1948).
\bibitem{b18}T. Balcerzak, Physica A\textbf{317},  213 (2003).
\bibitem{b19}A. Scherz, C. Sorg, M. Bernien, N. Ponpandian, K. Baberschke,
H. Wende, P. J. Jensen, Phys. Rev. B \textbf{72},  054447 (2005).
\bibitem{b20}A. Ney, F. Wilhelm, M. Farle, P. Poulopoulos, P. Srivastava, K.
Babrschke, Phys. Rev. B \textbf{59},  R3938 (1999).
\bibitem{b21} F. May, P. Srivastava, M. Farle, U. Bovensiepen, H. Wende, R. Chauvistre, K.
Babrschke, J. Magn. Magn. Mater. \textbf{177-181},  1220 (1998).
\bibitem{b22}U. Bovensiepen, F. Wilhelm, P. Srivastava, P. Poulopoulos, M. Farle,
A. Ney, K. Baberschke, Phys. Rev. Lett. \textbf{81},  2368 (1998).
\bibitem{b23}P. J. Jensen, K. H. Bennemann,  P. Poulopoulos, M. Farle, F. Wilhelm,
K. Baberschke, Phys. Rev. B \textbf{60},  R14994 (1999).
\bibitem{b24}A. Scherz,
F. Wilhelm, P. Poulopoulos, H. Wende, K. Baberschke, J.
Synchrotron Rad. \textbf{8},  472 (2001).
\bibitem{b25}Ney A, Poulopoulos P and
Baberschke K  Europhys. Lett. \textbf{54},  820 (2001).
\bibitem{b26}M. A. Gusmao,
C. Scherer, phys. stat. sol.  (b) \textbf{92},  595 (1979).
\bibitem{b27}M. Zheng,
L. Menon, H. Zeng, Y. Liu, S. Bandyopadhyay, R. D. Kirby, D. J.
Sellmyer, Phys. Rev. B \textbf{62}, 12282 (2000).
\bibitem{b28}L. Wojtczak, J.
de Phys.  \textbf{30},  578 (1969).
\bibitem{b29}J. Pearson, Phys. Rev. A \textbf{138},
 213 (1965).
\bibitem{b30} W. Brodhorb, W. Haubeureisser, phys. stat.
sol.  (b) \textbf{8},  K21 (1965).
\bibitem{b31}L. Valenta, W. Brodhorb, W.
Haubeureisser, phys. stat. sol.  (b) \textbf{26},  191 (1968).
\bibitem{b32}B.
Busiakiewicz, I. Zasada, Phys. Rev. B \textbf{78}, 165412 (2008).


\end{thebibliography}
\end{document}